# A project of the thermal triple axis spectrometer at the PIK reactor

M.H. Iuzviuk*, N.A. Kovalenko, V.S. Tolichev, A.K. Ovsianikov, O.V. Usmanov, A.N. Matveeva, I.A. Zobkalo, V.V. Tarnavich.
*Petersburg Nuclear Physics Institute named by B.P.Konstantinov of NRC «Kurchatov Institute», Orlova Roshcha 1, Gatchina, 188300, Russia*

(*Electronic mail: yuzvyuk_mh@pnpi.nrcki.ru.)

## Abstract

A project of the novel thermal triple axis spectrometer IN1 at the PIK reactor is presented. Monte Carlo simulations were used for optimization its main parameters. The optimization of the instrument gives an estimated neutron flux as high as $10^9$ n/cm$^2$/s at the sample position at neutron energy about 36 meV at the designed 100 MW power. The modelling of neutron stages like monochromators, analyzers, virtual source, as well as instrument shielding was fulfilled to obtain the best performance of the spectrometer.

## Introduction

From the invention of the three-axis spectrometer (TAS) in the fifties of the previous century, this kind of neutron experimental technique remains the effective and highly demanded method for the exploring condensed matter. The flexible configuration of TAS allows obtaining an inelastic neutron response at any specified point in the energy-momentum space. Despite the successful development of time-of-flight (TOF) spectrometers for spallation neutron sources, instruments that allow for the rapid scanning of vast regions of reciprocal space, triple-axis spectrometers retain their importance. This is due to their unique ability to perform precision studies of localized regions of reciprocal space, which offers greater flexibility in experimental approach. Inelastic neutron scattering method is effective for studying dynamic processes in solid-state systems and provides detailed information about atomic and magnetic dynamics within the material (Figure 1). By analyzing the inelastic scattering spectrum, it is possible to extract the dispersion curves of phonons [1-3], which allows studying the interatomic interactions, critical phenomena, diffusion processes, anharmonicity of vibrations that govern heat capacity properties. This method is also considered to be the only direct way to the study of spin waves (magnons) in magnetically ordered substances, providing information on magnetic exchange energies, anisotropy, crystal-field splitting, and magnetic frustrated systems or spin-liquid states [4-6]. Dynamic properties play an important role across condensed matter physics, chemistry, and materials engineering, since macroscopic processes occurring within materials are intrinsically connected to these dynamic characteristics.

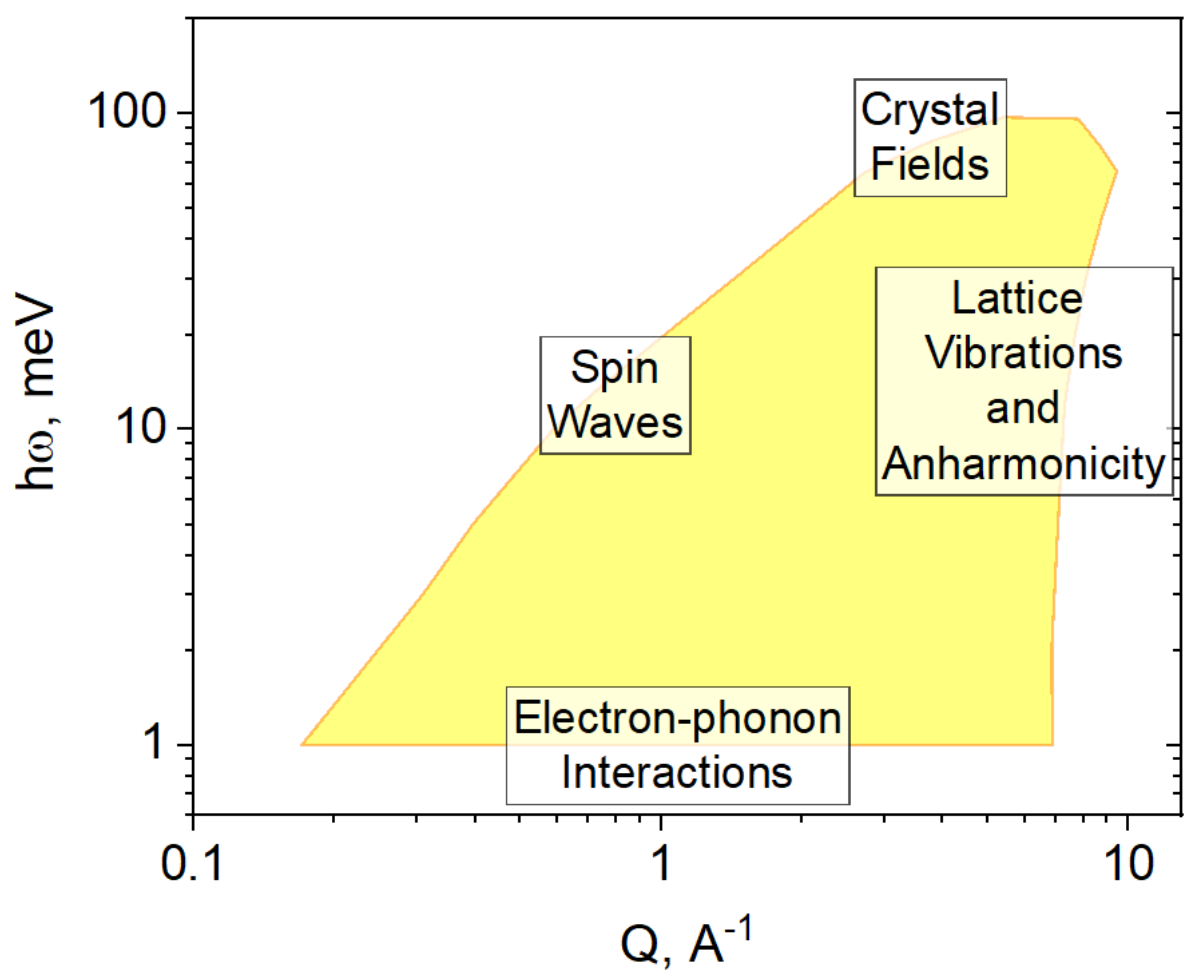

*Figure 1. The estimated dynamic range of IN1 (based on planned characteristics and [7]).*

This paper describes the design and the results of computer modeling of the parameters of neutron thermal three-axis spectrometer IN1. The calculations were performed using Monte Carlo (MC) simulation in the software package McStas [8, 9] and PHITS software package [10]. The modelling of IN1 spectrometer was performed taking into account its location in the reactor hall of the reactor PIK. Reactor PIK is a new high-flux research nuclear reactor at Petersburg Nuclear Physics Institute named by B.P. Konstantinov of National Research Centre «Kurchatov Institute» (Russia, Gatchina). The PIK reactor possesses a number of unique characteristics. These include, in particular, higher neutron flux density in the reflector zone, a highly efficient neutron trap with an exceptionally intense flux, and the ability to conduct radiation testing of materials directly in the reactor core. A key parameter is the maximum unperturbed thermal neutron flux, reaching $5\times10^{15}$ n/cm$^2$/s [11]. Reactor PIK has reached megawatt power in 2022 [12] and this first stage of operation at megawatt power confirms the achievability of the design parameters. Spectrometer IN1 will be created at the main reactor hall on the channel HEC10 (Figure 2), the expected flux of thermal neutrons at the exit of that channel is estimated as $2\times10^{11}$ n/cm$^2$/s.

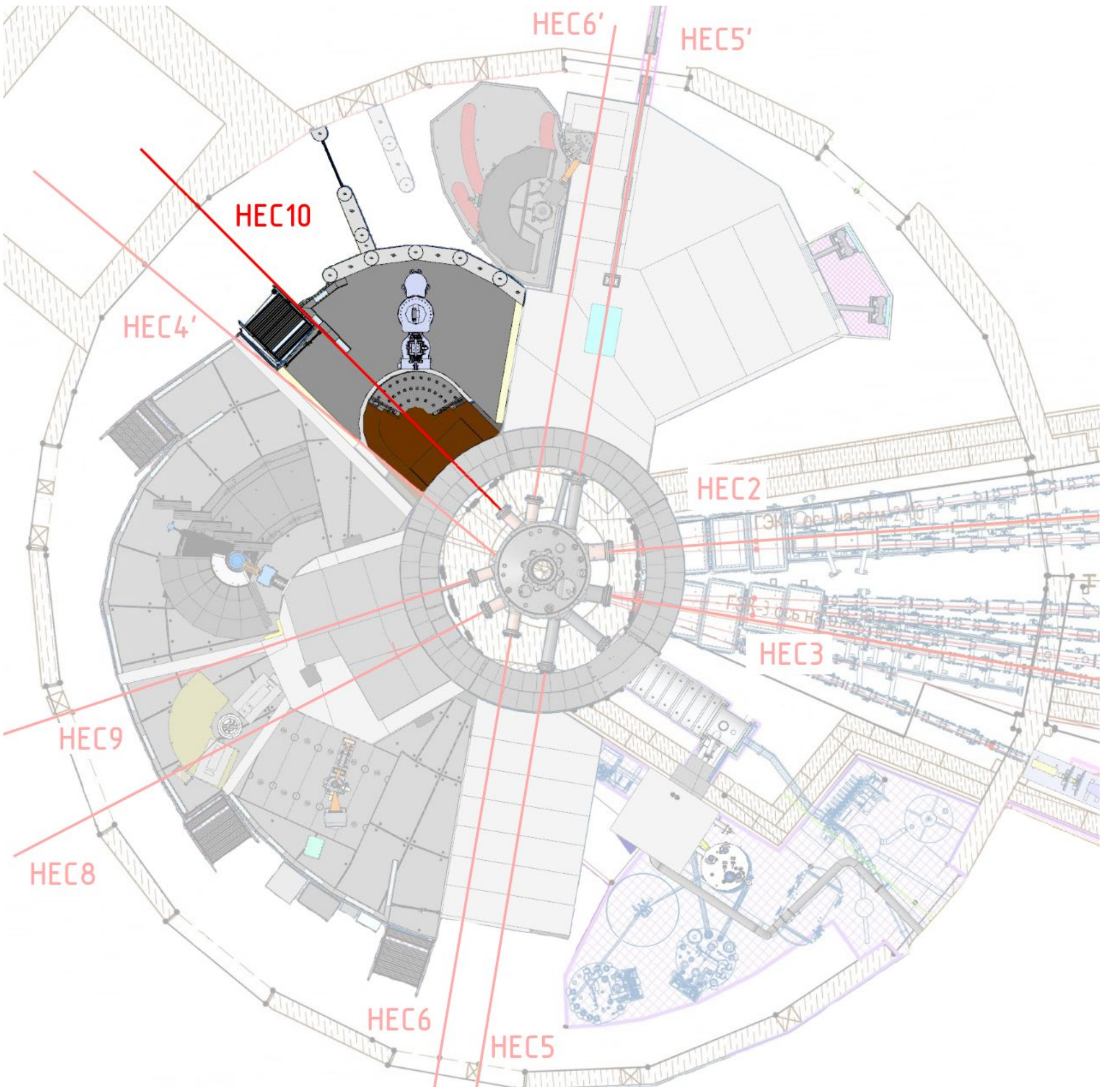

*Figure 2. The hall of Horizontal experimental channels of PIK reactor.*

IN1 is the first spectrometer in a series of inelastic instruments at PIK reactor, which includes: IN1 – thermal neutron spectrometer, IN2 – cold neutron spectrometer with polarization option, and IN3 – polarized thermal neutron spectrometer. This spectrometer complex will allow performing a broad scope of researches at the large energy transfer range using unpolarized as well as polarized neutrons. The position of the instrument IN1 on a channel with a high neutron flux, the use of a NICOS-based control routine with user-friendly interface [13], and machine learning methods [14] will significantly reduce the time needed to perform standard inelastic experiments.

## Instrument project description

Conventional scheme of three-axis spectrometer with accepted now functional units' arrangement is in the base of spectrometer IN1 project. The numerical simulations were undertaken in order to determine the optimal characteristics of spectrometer's units. First of all, it was necessary to decide how to set up a virtual source, which is necessary for the Rowland geometry implementation. From the first estimates it was shown that diaphragm with flexible changing slit cannot be effective in this case because of vicinity to the very intensive source. For its flexibility, such diaphragm should have rather light shutters, which obviously could not be effective against high flux. Therefore, the virtual source was chosen as massive steel block with permanent slit. Neutron beam goes through the virtual source to selected one of three monochromators, which considered to be based on: HOPG(002), pressed Ge(220), elastically curved Si(111). After that, neutrons of chosen energy scatter on a sample and then their energy is analyzed using one of two analyzers: either based on HOPG(002) or Si(111). Detection is performed using a single $^3$He-based detector. In order to choose necessary beam divergence, it is planned to use Soller collimators with divergence of 20′, 40′, 60′, which will be installed in front of the sample – it will be inserted into the mask – and in front of the analyzer, inserted into its shielding. Beside this, IN1 has two diaphragms, which serve to form the neutron beam and two low-efficiency neutron counters - monitors, which are necessary for control of neutron beam intensity. These units will be located before and after the sample stage.

Shielding of monochromator and analyzer will consist of a set of mobile sectors in order to ensure the smooth changing of neutron energy, incident on the sample. The open sector corresponds to the specified neutron energy, which is selected by the monochromator from the thermal neutron spectrum.

The schematic layout of instrument and its main characteristics are presented in Figure 3 and Table 1. Horizontal channel HEC10 has 215 cm height above the floor in the PIK main reactor hall. Therefore, IN1 tanzboden will be placed on the platform in order to provide satisfactory height of the spectrometer units and convenience of the manipulation with the sample and sample environment.

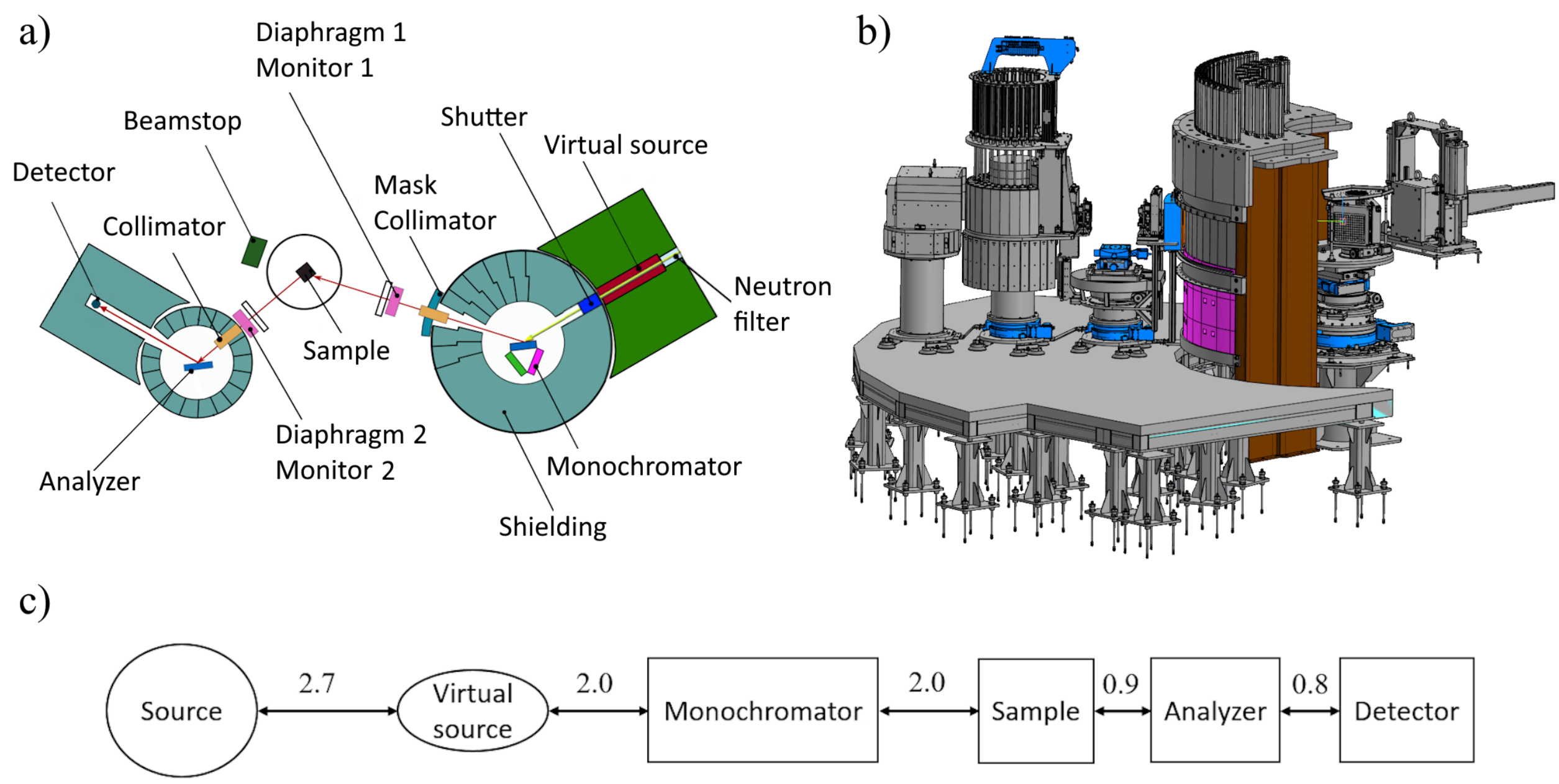

*Figure 3. Scheme (a) and 3D model (b) of thermal triple axis spectrometer IN1. Stationary monochromator shielding is not shown. c) The distance between the main points of instrument is given in meters.*

*Table 1 – The main planned characteristics of thermal triple axis spectrometer IN1*

| | |
|---|---|
| Location | Reactor hall, thermal channel HEC10 |
| Beam crossection (H×W) | 20×6 cm |
| Source to monochromator distance | 4.70 m |
| Virtual source to monochromator distance | 2 m |
| Monochromator to sample distance | 2 m |
| Sample to analyzer distance | 0.9 m |
| Analyzer to detector distance | 0.8 m |
| Neutron energy range | 15 meV < $E_i$ < 100 meV |
| Resolution (FWHM) | $\Delta E$ = 0.8 meV ($E_i$ = 15 meV)<br>$\Delta E$ = 3 meV ($E_i$ = 35 meV) |
| Double focusing monochromators | PG(002), Ge(220), Si(111)<br>10° < $2\theta_M$ < 90° |
| Double focusing analyzers | PG(002), Si(111)<br>–110° < $2\theta_A$ < 110° |
| Sample angle | –90° < $2\theta_S$ < 90° |
| Detector | Single $^3$He |
| Soller collimators (optional) | 20′, 40′, 60′ |

## Monte Carlo simulations

Monte Carlo simulations were carried out to get optimal sizes of input slit, monochromators, analyzers as well as instrument shielding.

### 1. Sapphire filter and virtual source (VS)

The IN1 spectrometer will use Rowland geometry to achieve optimal focusing conditions. It means that the source, the monochromator and the sample lie on one circle and monochromator–source distance and monochromator–sample distance are equal to each other. To ensure these conditions, the concept of a virtual slit-shaped source is used, which has been successfully implemented in a number of modern triple axis spectrometers [15-19]. The combined use of the virtual source and the sapphire filter significantly reduces parasitic neutron and gamma radiation. The calculated neutron spectrum obtained using the sapphire filter is shown in Figure 4.

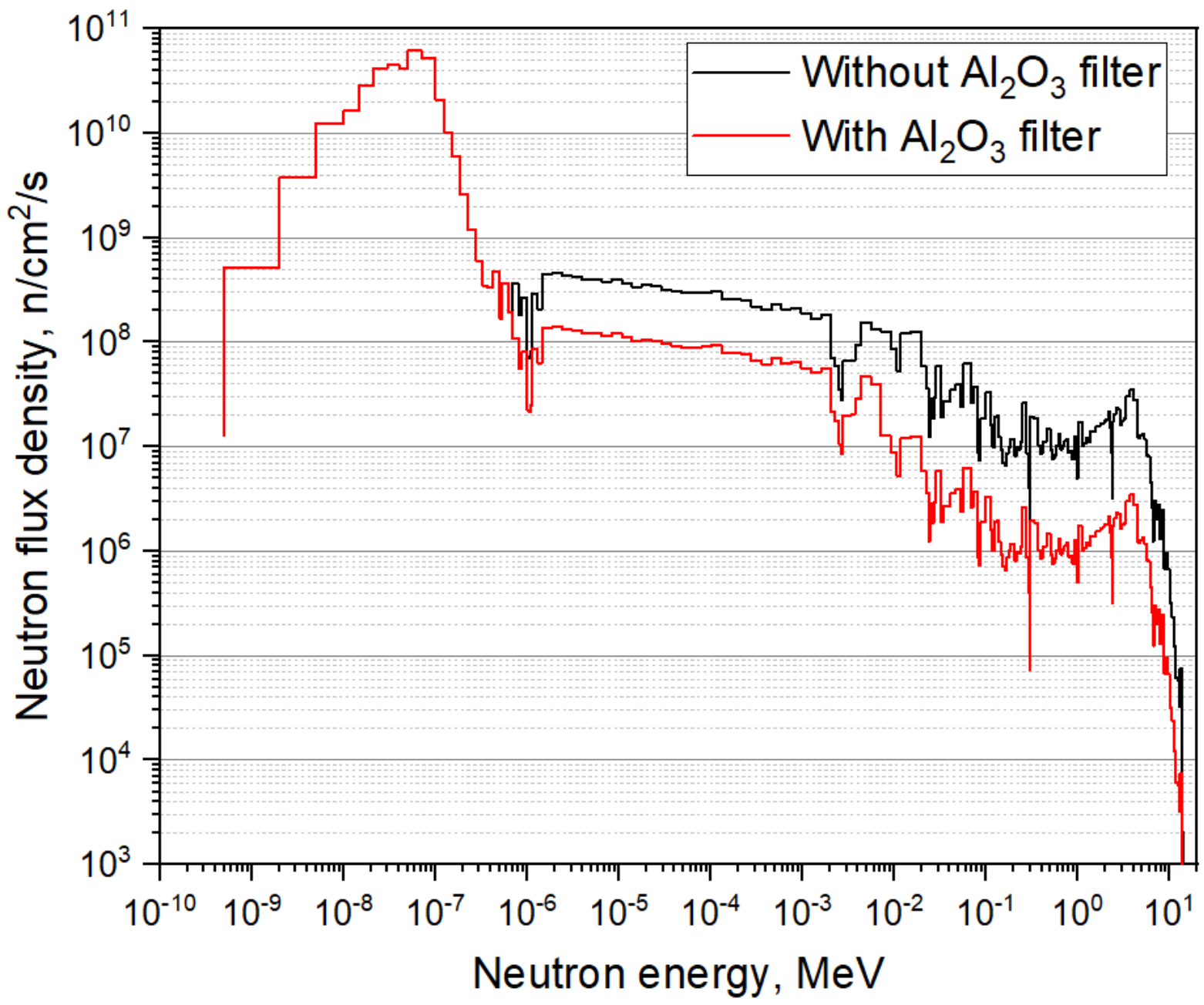

*Figure 4. Neutron spectrum at the exit of a VS with a sapphire filter*

Since the parameters of the VS slit are fixed and must remain unchanged, the optimal characteristics of the VS were determined taking into account its efficiency across a wide range of energies. The relative change in neutron flux density for various wavelengths as a function of the virtual source width and height is shown in Figure 5.

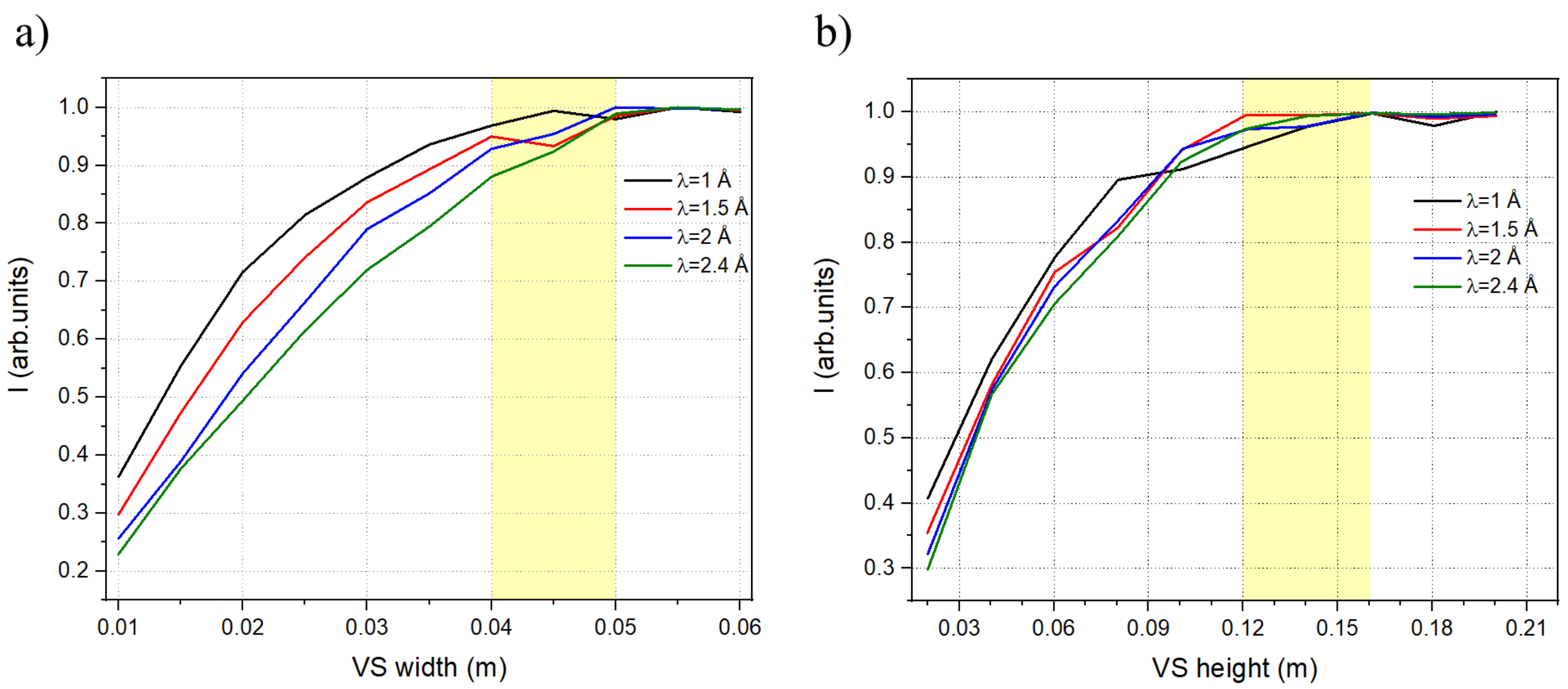

*Figure 5. Relative change in neutron flux density for different wavelengths depending on the width and height of the virtual source. The colored areas show the optimal width and height.*

As it can be seen from the figure 5, the optimal width is between 4 and 5 cm, and the height is between 12 and 16 cm. Therefore, following dimensions of the virtual source (W×H×L) were adopted: 4×12×40 cm$^2$. The length of VS was chosen after calculations of background at the sample position. With such virtual source dimensions, the background level at the sample is minimized (59% lower than the background from the source), while the flux of useful neutrons at the sample does not decrease appreciably.

Figure 6 shows the wave vector dependence of neutron intensity and Figure 7 shows calculated beam profile at the distance of 10 cm in front of the monochromator without VS (a) and with it (b).

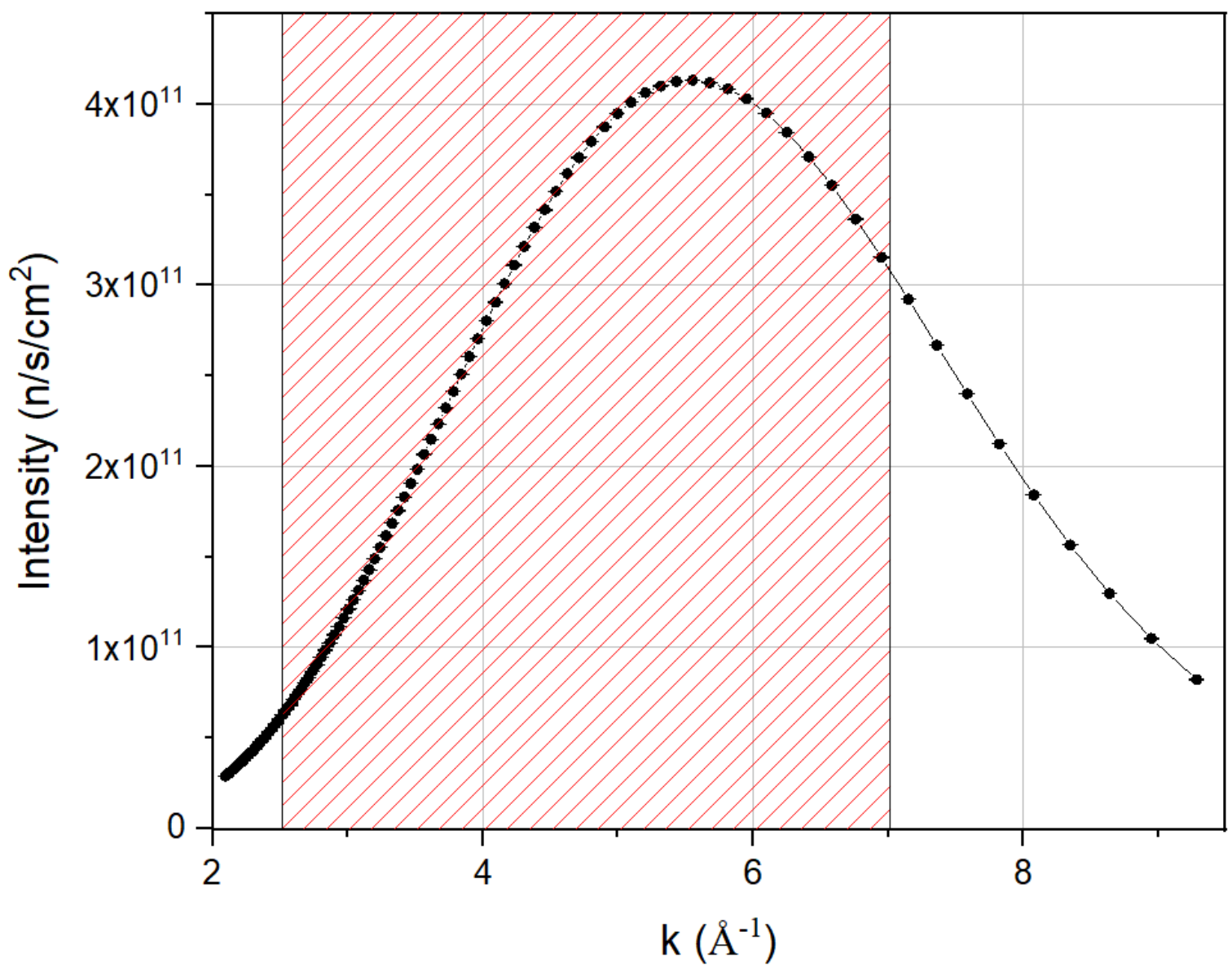

*Figure 6. Neutron intensity vs. wave vector k. The red area corresponds to working range of k values of IN1.*

a) b)

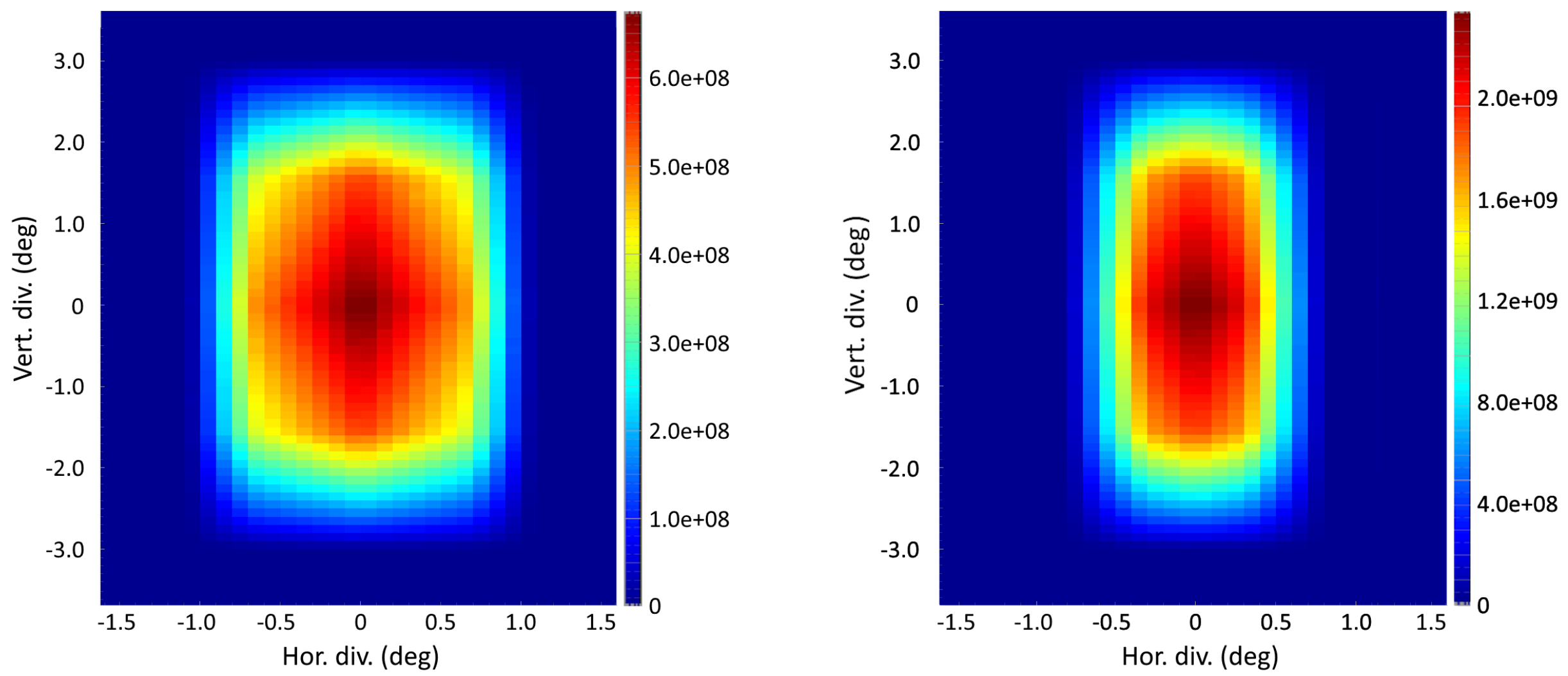

*Figure 7. Beam profile in front of monochromator unit a) without VS and b) with VS.*

## 2. Monochromators

IN1 spectrometer will be equipped with three monochromators based on: HOPG (002), pressed Ge (220) and elastically bent Si (111). The first one is chosen as monochromator with the best reflectivity for neutrons and will be used in the standard configuration, Ge-based monochromator is intended to operate at higher energy transfer, and elastically bent silicon monochromator allows to achieve high resolution. The total size of monochromator working area and size of single crystals were chosen by taking in account the following considerations:

1. Maximum neutron flux intensity at sample position.
2. The odd number of rows and columns.
3. Unification of size of HOPG monochromator and Ge one.
4. Geometry restrictions.

In order to optimize the neutron flux MC simulations were carried out, the results are shown in Figure 8 and 9. For the calculations the mosaic spread of 48′ for HOPG and 30′ for Ge were taken. In this way the optimal values for dimensions of monochromator working area and crystals size were obtained. Thus, total working area obtained to be equal 26×22 $cm^2$ (W×H), with HOPG crystals size equal to 20×20×2 $mm^3$, Ge crystal size equal to 20×20×8 $mm^3$ (W×H×Th). To ensure the obtained parameters, each of the HOPG and Ge-based monochromators must consist of 143 single crystals.

The third monochromator consists of 7 composite silicon plates (250×32 $mm^2$), each of them made of ten 0.6 mm thick silicon crystals and located in holders of a vertically focusing device. The mosaic is introduced by bending plates in horizontal plane.

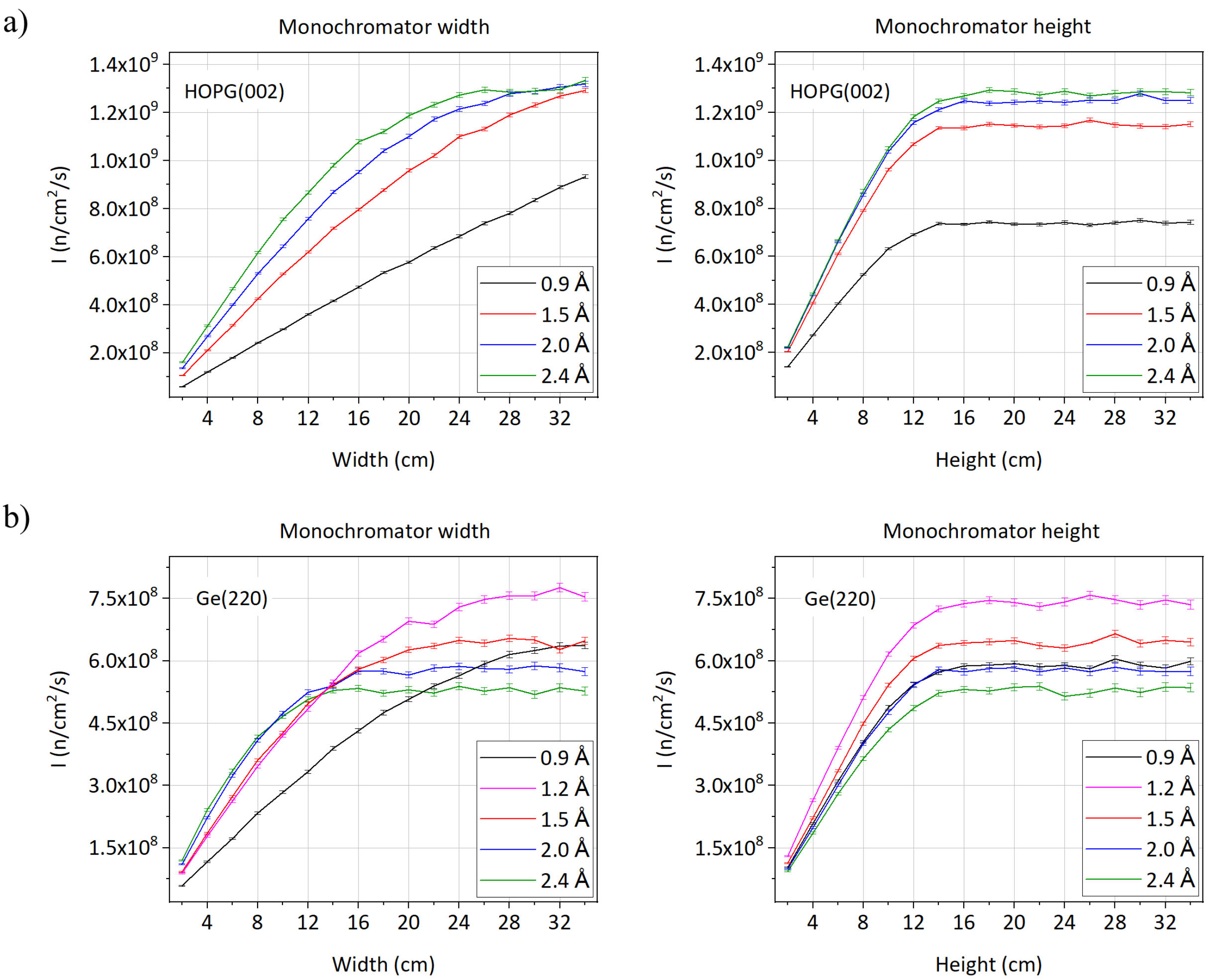

*Figure 8. Simulations of the neutron flux intensity at the sample position using a) PG monochromator with various width and height, b) Ge monochromator with various width and height.*

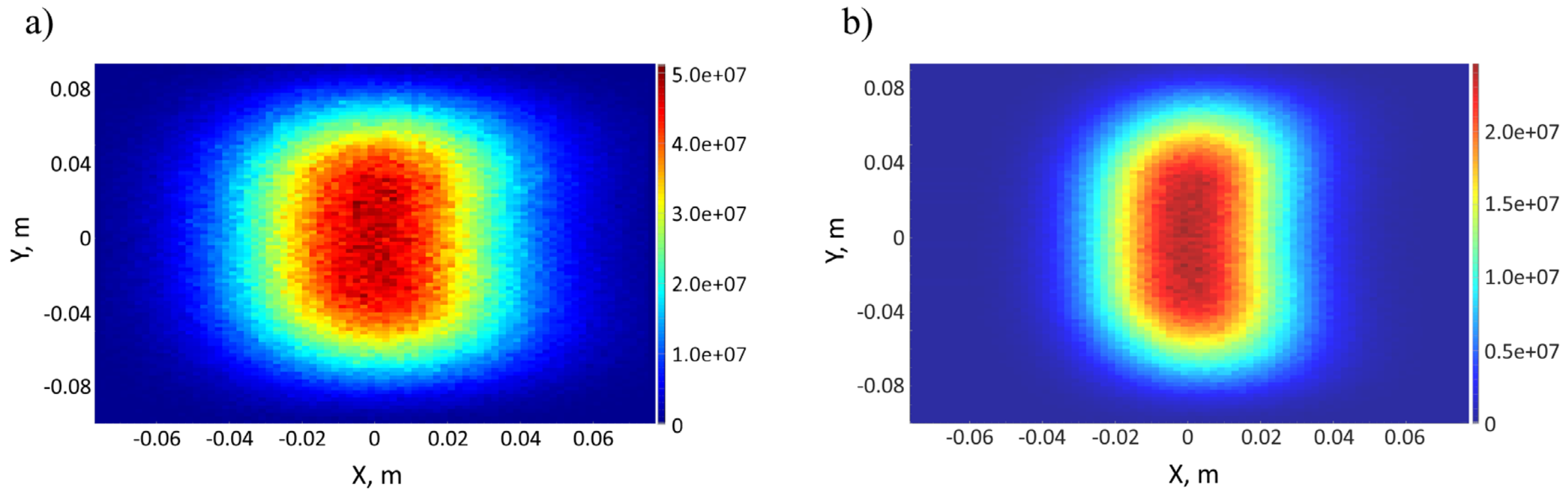

*Figure 9. a) Neutron beam profile at the sample position from HOPG monochromator, λ=1.5Å, b) Neutron beam profile at the sample position from Ge monochromator, λ=0.9Å.*

### 3. Analyzers

IN1 analyzer unit will be equipped with two monochromatizing devices: one based on HOPG and another based on Si crystals, the use of which will depend on the monochromator being used. The simulations for optimizing the neutron flux on the detector after analyzers were carried out in the same way as in the case of monochromators. Results of the simulations with the use of HOPG monochromatizing unit and slit as sample presented at Figure 10. The obtained optimal size of working area of HOPG analyzer is equal to 22×14 $cm^2$ (W×H), it consists of 77 single crystals with dimensions 20×20×2 $mm^3$. The design and parameters of the silicon analyzer are similar to those of the silicon monochromator, except that it contains 5 silicon plates.

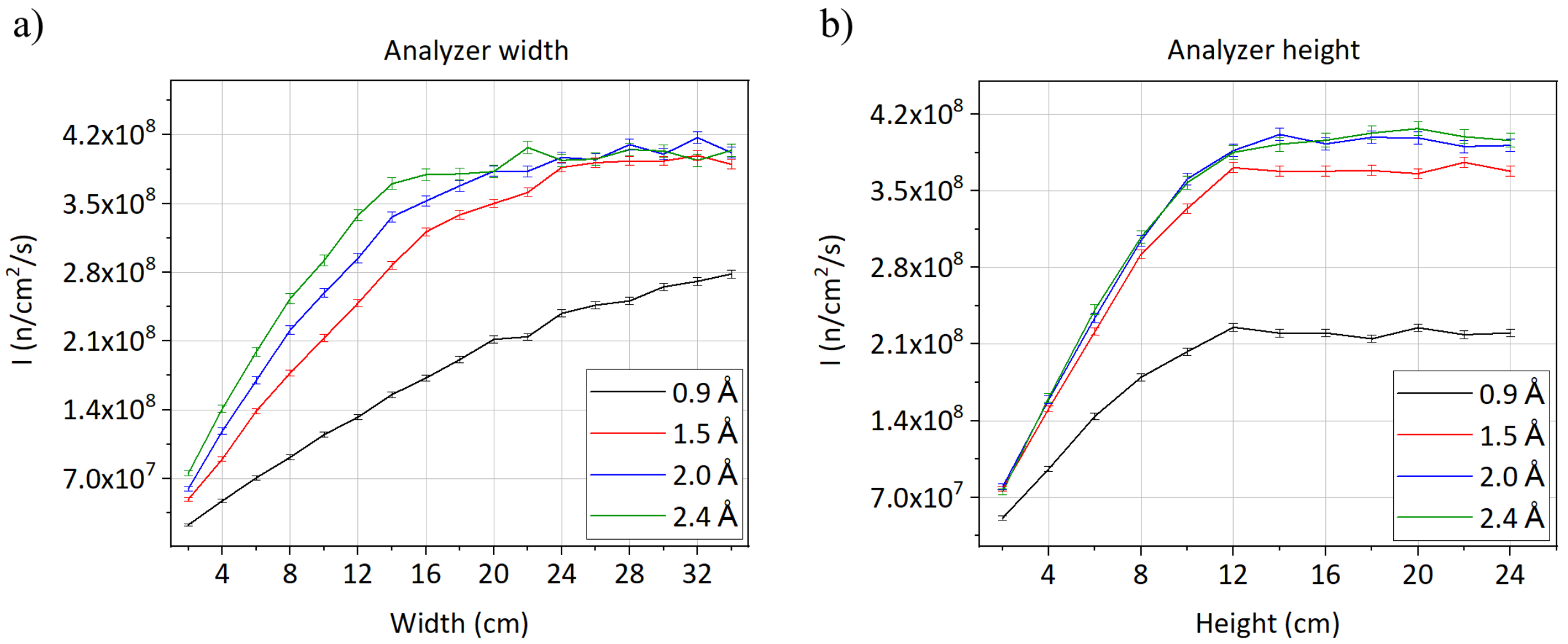

*Figure 10. Simulation of the neutron intensity at the detector position using PG analyzer with various width (a) and height (b).*

## 4. Instrument shielding

To simulate the radiation spatial dose distribution, a mathematical model was developed using the PHITS software package [10]. A general view of the spectrometer model, which were used for radiation dose calculations is shown in Figure 11.

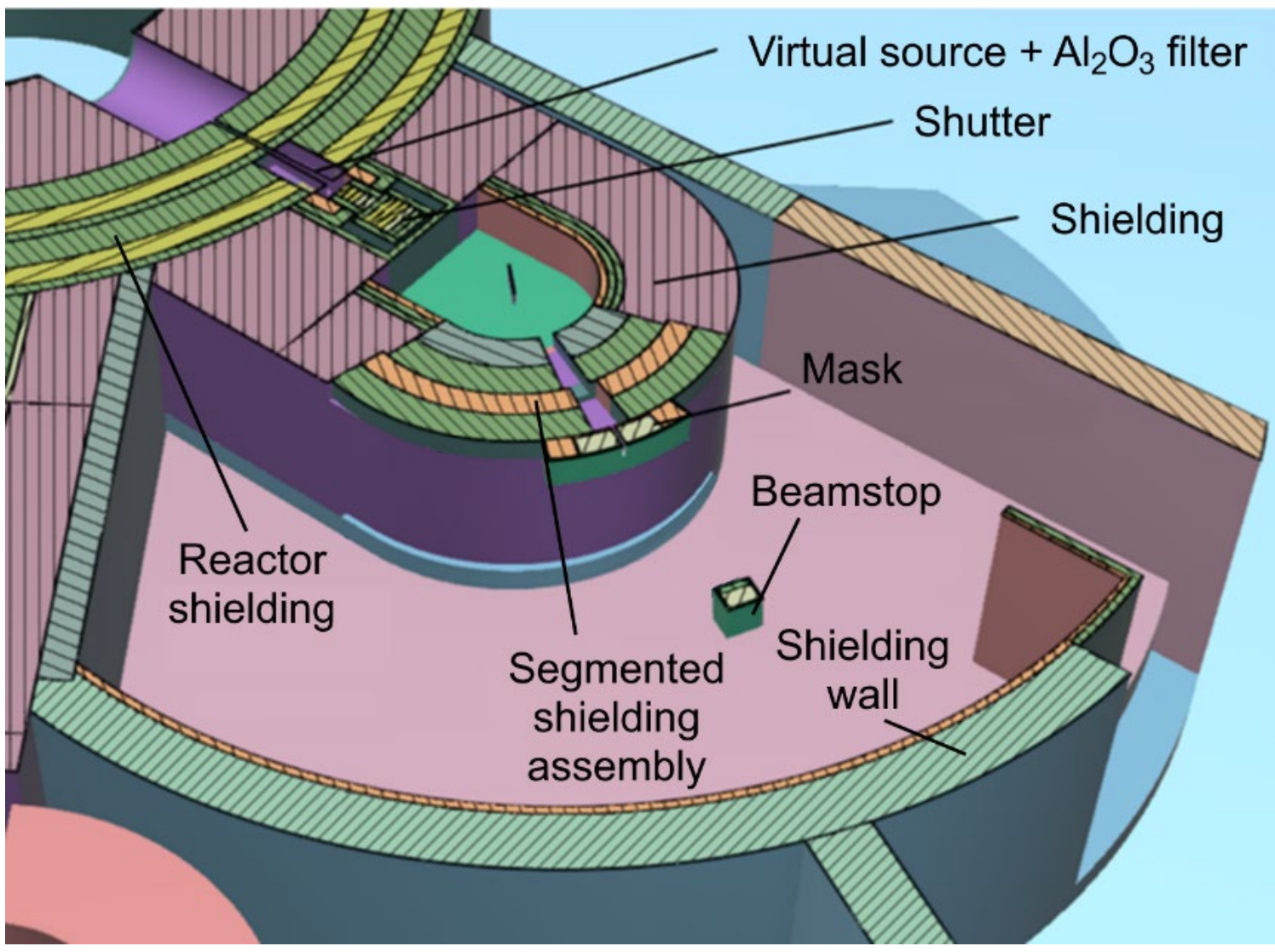

*Figure 11. The general view of the model of IN1, used for radiation dose calculation. Section in the horizontal plane at the beam level.*

### 4.1. Shutter

The shutter is designed to block the neutron beam during some short activity at the spectrometer working area like sample changes, sample environment installation etc. (or in emergency situations). During an experiment, the shutter is raised so that the neutron beam passes unimpeded to the monochromator unit. It consists of 15 cm of steel, 30 cm of borated polyethylene (BP) and 15 cm of lead.

### 4.2. Fixed monochromator shielding

Fixed monochromator shielding (Figure 12) consists of 5 cm of borated polyethylene BP-5 (red), 10 cm of steel (dark green) and 71 cm of heavy concrete with $\rho = 3.6$ g/cm$^3$ (purple). Directly above the monochromator is the monochromator protective cover, designed as a separate unit. That design solution provides access to the monochromator components, when necessary, without having to completely disassemble the protective housing. This unit consists of a 5 cm thick layer of PB-5, a 20 cm thick steel layer, and a 60 cm thick layer of heavy concrete.

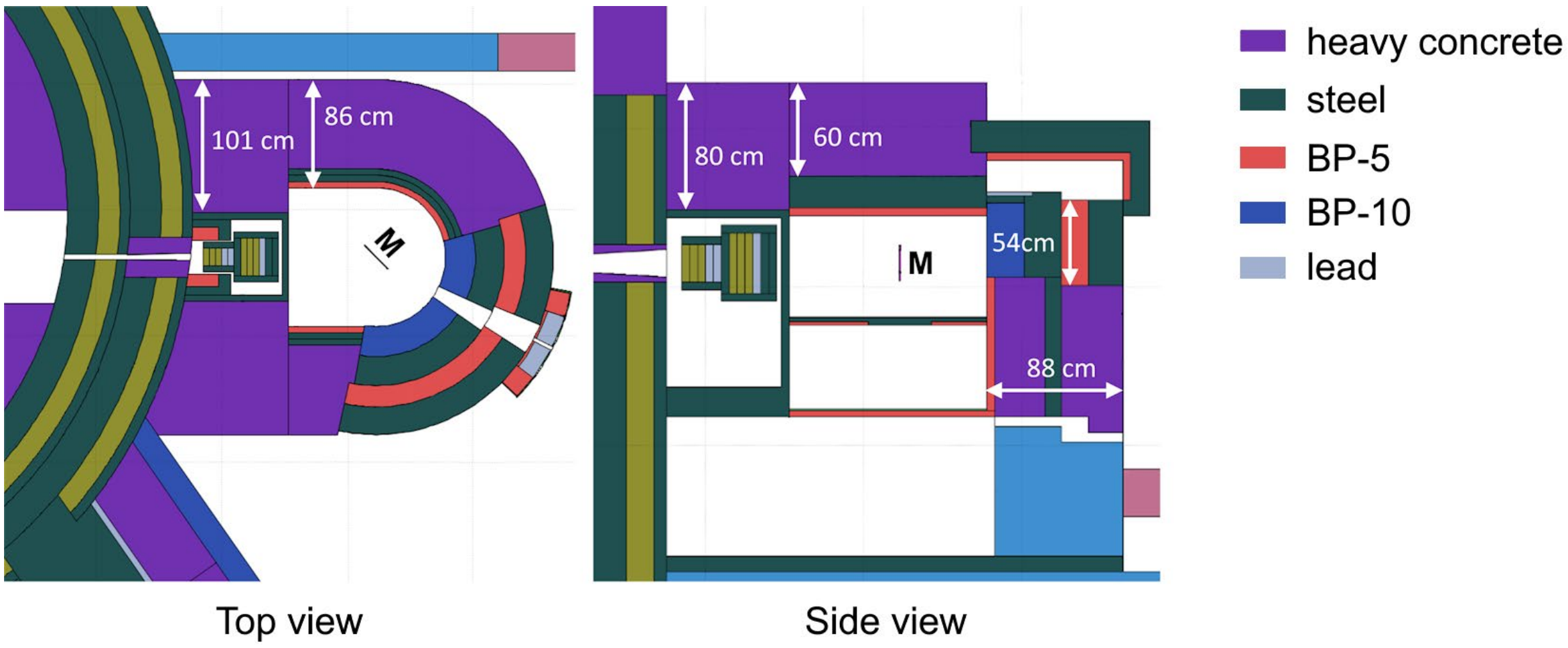

*Figure 12. General view of the monochromator shielding design*

The results of dose rate modeling for the open-shutter mode at the beam level are shown at Figure 13.

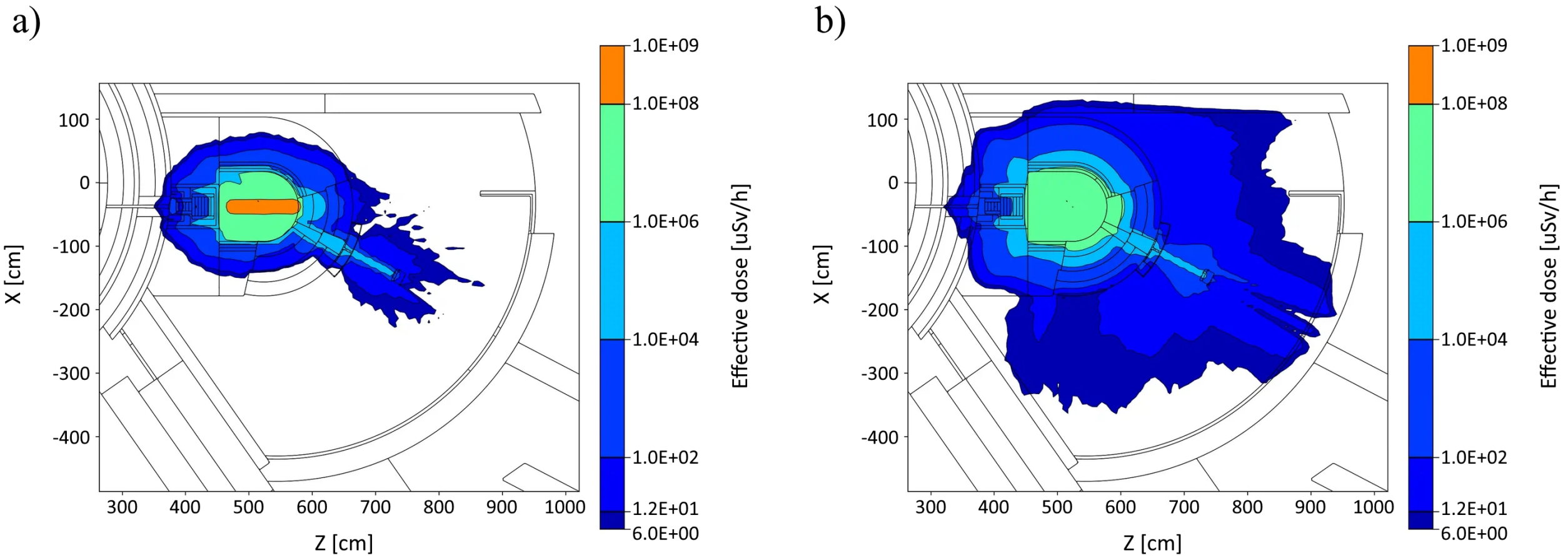

*Figure 13. Spatial distribution of a) neutron dose rate, b) gamma-ray dose rate, at the spectrometer working area.*

### 4.3. Segmented shielding assembly

The segmented monochromator shielding is a movable part of the overall monochromator's shielding. Its design provides a neutron access angle to sample position up to 105° and consists of 5°-wide liftable sectors. During an experiment, a sector (or set of sectors) corresponding to the required neutron energy is lifted, allowing the beam to pass through. This ensures smooth adjustment of the monochromator to a given initial neutron energy in the range from 15 meV to 100 meV, and neutrons with the corresponding energy to be delivered to the sample.

In the case of IN1, the segmented shielding assembly consists of two rows of sectors: the inner row consists of 24 cm of PB-10 (blue) and 23 m of steel (dark green), the second row consists of 17 cm of PB-5 (red) and 22 cm of steel (dark green) (Figure 12 top view). The first row also contains thin layer of lead (2 cm) at the top of sectors. Segmented shielding is placed on a base made of heavy concrete (purple), which provides additional radiation shielding and mechanical stability of the system (Figure 12 side view).

Modeling of the monochromator shielding revealed a vulnerable point, designated as * in Figure 14, through which radiation from the monochromator's mechanical components could penetrate. To reduce radiation levels in the spectrometer's operating area, an additional shielding was designed and positioned above the segmented one (as shown in Figure 14). This shielding consists of steel elements (dark green) of varying thicknesses and a 5 cm thick layer of PB-5 material (red). This configuration reduces the dose rate at the shielding's outer surface to 10 µSv/h, thereby significantly improving background conditions in the spectrometer's operating sector and eliminating adverse effects on personnel working with other instruments in the reactor hall.

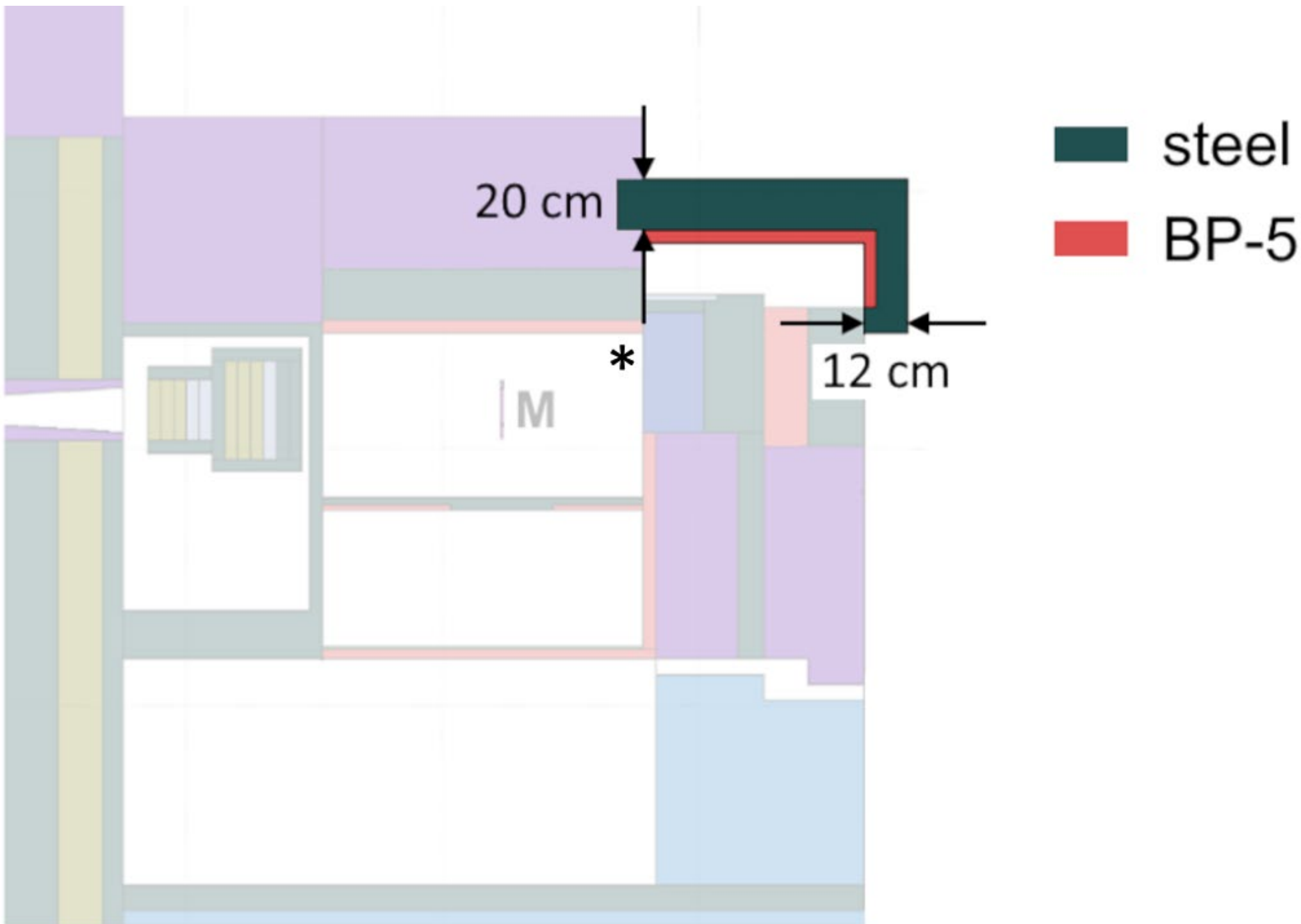

*Figure 14. General view of the additional shielding construction above the sectors.*

### 4.4. Mask

A shielding mask is designed to block possible trajectories of background neutrons reflected from the monochromator structures through the open segment. Mask is made of PB-5, lead and a layer of boron rubber with 60% of boron, is installed in the area adjacent to the segmented shielding assembly. All calculated dimensions are shown in Figure 15.

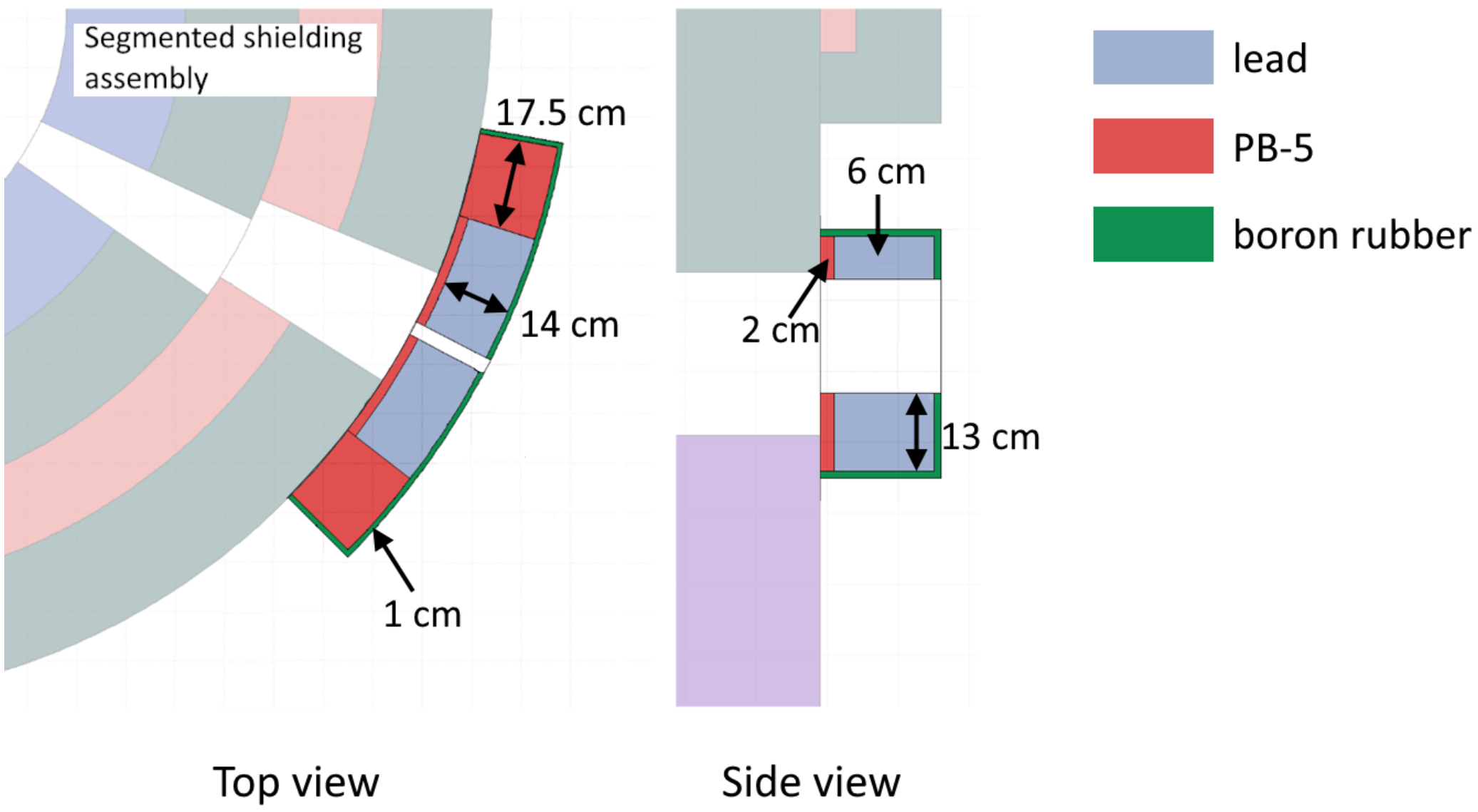

*Figure 15. General view of the shielding mask.*

### 4.5. Beamstop

The beamstop is designed to absorb neutrons reflected from the monochromator toward the sample without being scattered or absorbed. It will be placed on the sample stage in front of incoming neutron beam. It consists of 4 cm of PB-10 and 12 cm of lead coated with a layer of borated rubber, thus its thickness is equal to 18 cm with transverse dimensions 22×46 cm. In the case when analyzer will take position close to the direct beam from the monochromator, beamstop goes down so as not to obscure useful reflections.

### 4.6. Shielding wall

Spectrometer working area is surrounded by a 35 cm concrete wall with a 5 cm PB-5 insert at the neutron beam level (Figure 16). This design ensures a minimal equivalent dose rate (DER) level outside the working area of the facility.

The simulations in the PHITS software package, visualized in Figure 16, shows that creating a shield around the monochromator's working area significantly reduces and localizes the propagation of neutron and photon DERs within the monochromator's shielding area. A concrete wall with a PB-5 insert reduces the DER behind the shielding to 1 μSv/h.

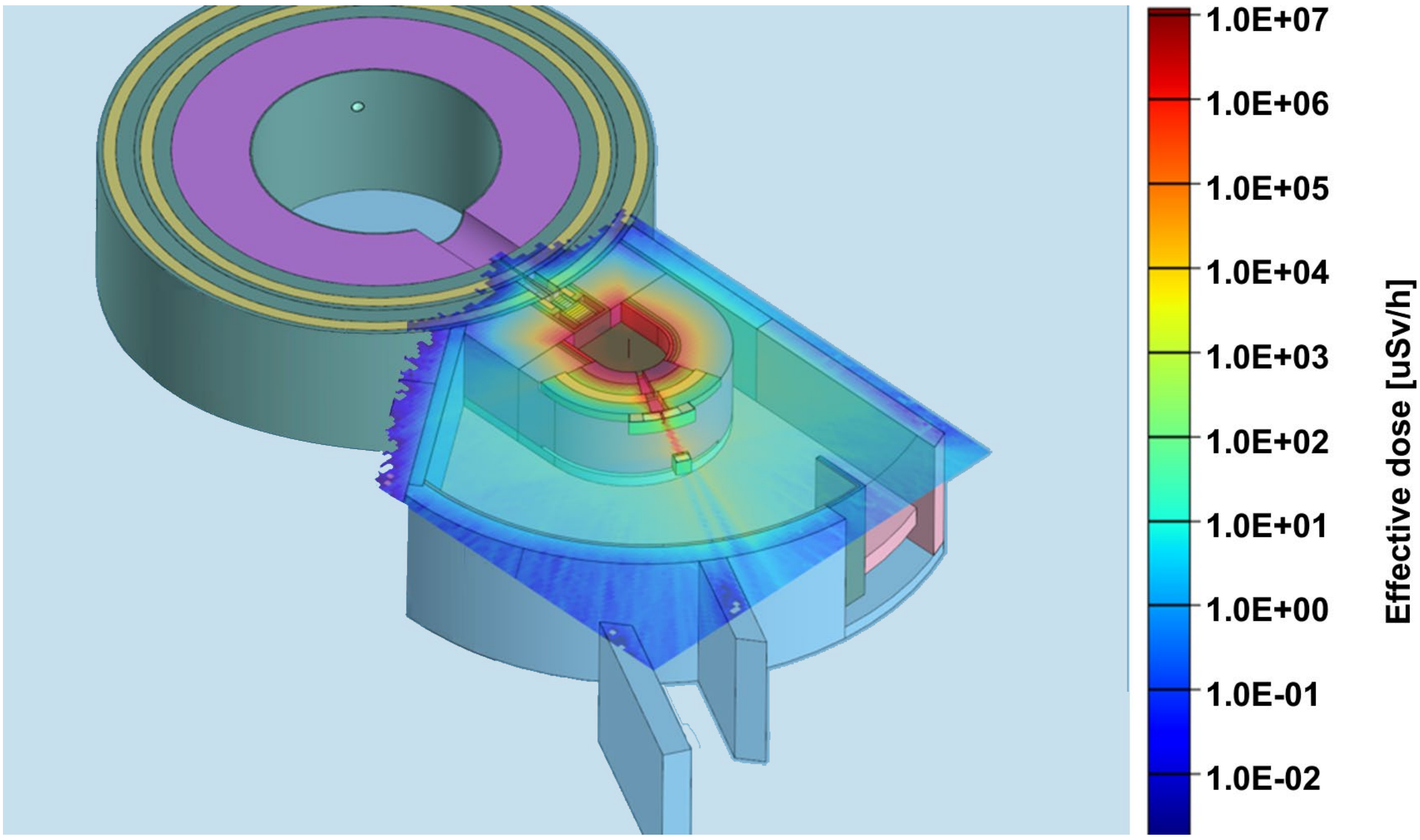

*Figure 16. Distribution of DER within shielding of IN1*

### 4.7. Shielding of analyzer and detector units

Shielding of an analyzer is necessary to reduce background radiation from γ-radiation, thermal neutrons, and fast neutrons on the analyzer. It must allow smooth adjustment of the analyzer's vertical plane rotation to a specified angle, i.e., smooth adjustment to the required neutron energy. For this purpose, a segmented shielding assembly is used, similar to that of the monochromator. The open segment corresponds to the analyzer's operating rotation angle, which, in turn, corresponds to the required neutron energy. Segment of analyzer shield consists of 6 cm PB-20 and 14 cm PB-10, its outer and inner sides are covered with a 5 mm thick layer of boron-contained rubber (Figure 17).

Shielding of the detector unit is necessary to reduce the background from γ-radiation, thermal and fast neutrons at the detector. It consists of several layers of different materials: the first inner layer is 2.5 mm of cadmium, then there are 5 mm of boron rubber, 6 cm of PB-20, 14 cm of PB-10, and 5 mm of boron rubber (Figure 17).

The simulations were carried out taking into account the geometric limitations associated with the analyzer and detector being located close to each other.

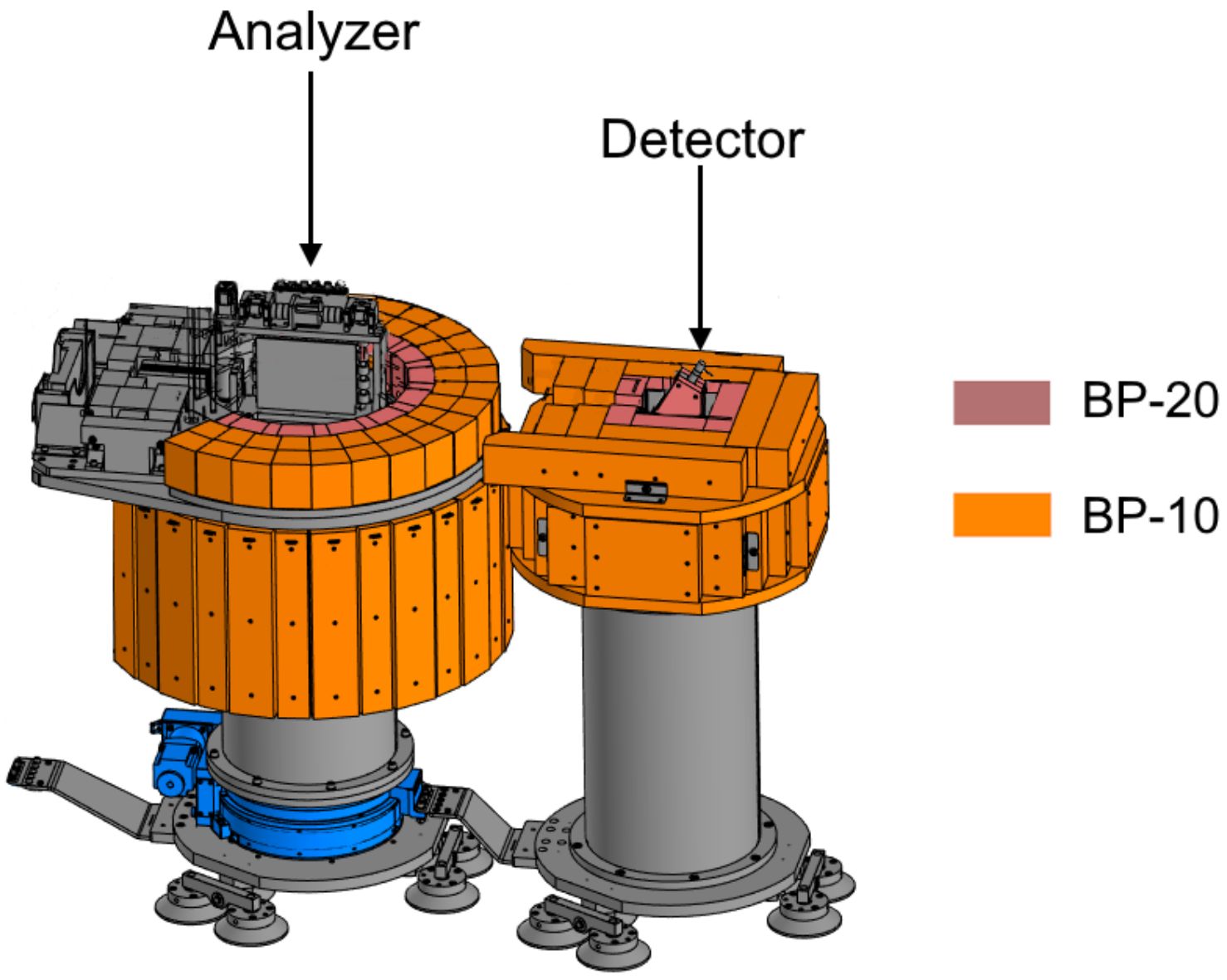

*Figure 17. Shielding of analyzer and detector. Upper part of the shielding is not shown. Boron rubber and cadmium are not shown.*

## Conclusions

This article presents the project of thermal triple-axis spectrometer IN1, which will be created at the new high-flux reactor PIK. MC optimization was carried out to obtain the best performance of the separate stages as well as of whole instrument. The performed modeling allowed us to determine materials for the radiation protection of the spectrometer units and the entire working sector. Specific geometric dimensions for all of them were obtained, enabling the required DER level to be achieved and the neutron background in the facility area to be reduced to a level safe for personnel servicing or located near the experimental setup as well as to minimize the background in the area of detector.

The expected maximum flux of neutrons at the sample position is about $1x10^9$ n/cm$^2$/s with the use of HOPG monochromator and about $6x10^8$ n/cm$^2$/s using Ge monochromator.

The use of a NICOS-based control routine with user-friendly interface [13], and machine learning methods [14] will significantly reduce the time needed to perform standard experiments.

Thus, IN1 can become a sought-after and useful “workhorse” for investigating inelastic phenomena within emerging material systems. Spectrometer IN1 is under construction now, more details about this instrument will follow in the next paper after commissioning.

## Author declarations section

The authors have no conflicts to disclose.

## Data availability statement

The data that support the findings of this study are available from the corresponding author upon reasonable request.